# Electronic structure of SrTi$_{1-x}$V$_x$O$_3$ films studied by *in situ* photoemission spectroscopy: Screening for a transparent electrode material

Tatsuhiko Kanda[1], Daisuke Shiga[1,2], Ryu Yukawa[2,†], Naoto Hasegawa[1], Duy Khanh Nguyen[1], Xianglin Cheng[1], Ryosuke Tokunaga[1], Miho Kitamura[2], Koji Horiba[2], Kohei Yoshimatsu[1], and Hiroshi Kumigashira[1,2,*]

[1] *Institute of Multidisciplinary Research for Advanced Materials (IMRAM), Tohoku University, Sendai, 980–8577, Japan*

[2] *Photon Factory, Institute of Materials Structure Science, High Energy Accelerator Research Organization (KEK), Tsukuba, 305–0801, Japan*

**Abstract**

This study investigated the electronic structure of SrTi$_{1-x}$V$_x$O$_3$ (STVO) thin films, which are solid solutions of strongly correlated transparent conductive oxide (TCO) SrVO$_3$ and oxide semiconductor SrTiO$_3$, using *in situ* photoemission spectroscopy. STVO is one of the most promising candidates for correlated-metal TCO because it has the capability of optimizing the performance of transparent electrodes by varying $x$. Systematic and significant spectral changes were found near the Fermi level ($E_F$) as a function of $x$, while the overall electronic structure of STVO is in good agreement with the prediction of band structure calculations. As $x$ decreases from 1.0, spectral weight transfer occurs from the coherent band near $E_F$ to the incoherent states (lower Hubbard band) around 1.0–1.5 eV. Simultaneously, a pseudogap is formed at $E_F$, indicating a significant reduction in quasiparticle spectral weight within close vicinity of $E_F$. This pseudogap seems to evolve into an energy gap at $x$ = 0.4, suggesting the occurrence of a composition-driven metal–insulator transition. From angle-resolved photoemission spectroscopic results, the carrier concentration $n$ changes proportionally as a function of $x$ in the metallic range of $x$ = 0.6–1.0. In contrast, the mass enhancement factor, which is proportional to the effective mass ($m^*$), does not change significantly with varying $x$. These results suggest that the key factor of $n/m^*$ in optimizing the performance of correlated-metal TCO is tuned by *x*, highlighting the potential of STVO to achieve the desired TCO performance in the metallic region.

[†] Present Address: *Department of Applied Physics, Osaka University, Suita, Osaka 565-0871, Japan*

[*]Author to whom correspondence should be addressed: kumigashira@tohoku.ac.jp

## I. INTRODUCTION

Transparent conducting oxides (TCOs) are of great technological interest for a myriad of applications, notably including flat panel displays, solar cells, touchscreen sensors, and other optoelectronic devices [1]. Transparent conduction in oxide semiconductors, such as Sn-doped $In_2O_3$ (ITO) and doped ZnO, is realized by degenerately doping electrons in the conduction band in a wide energy gap above visible light (3.25 eV) to realize an optimal balance between the two conflicting properties of optical transparency and electrical conductivity [1–4]. The optical transparency in low-energy regions is dominated by the free-carrier reflection edge, which is represented by the screened plasma energy $\hbar\omega_p = \hbar(e/\sqrt{\varepsilon_0\varepsilon_r})\sqrt{n/m^*}$, where $e$, $\varepsilon_0$, and $\varepsilon_r$ are the elemental charge, vacuum permittivity, and relative permittivity, respectively, and $n$ and $m^*$ are the free carrier concentration and effective carrier mass, respectively. The reflection edge must be minimized by choosing an appropriate ratio of $n$ to $m^*$. In contrast to the optical transparency, the electrical conductivity $\sigma = e^2\tau(n/m^*)$ with scattering time $\tau$ must be maximized. This fundamental principle makes the ratio $n/m^*$ the key factor in optimizing the performance of transparent conducting materials: the ratio must be maximized to enhance electrical conductivity but limited to keep the free carrier reflection edge below the visible light region. Doping the semiconductors increases $n$ while maintaining $m^*$. However, the highest $n$ achieved in semiconductors is below $3 \times 10^{21}$ cm$^{-3}$ owing to the solubility limit of dopants and pronounced self-compensation [5], although the resultant $\hbar\omega_p$ is well below the low-energy onset of the visible spectrum. This limit has prevented performance improvements in TCOs based on wide-gap semiconductors [3–6].

Recently, a new approach for TCOs has been proposed [7], taking advantage of the strong electron–electron interactions in highly correlated transition-metal oxides [7–15], such as "$n$-type" $SrVO_3$, $CaVO_3$ [7–11], and $SrNbO_3$ [12] as well as "$p$-type" $La_{1-x}Sr_xVO_3$ [13] and $V_2O_3$ [14]. These oxides essentially have metallic ground states, but the energies of interband optical transitions (O 2$p$ → transition metal $d$ band) primarily start from the blue region of the visible spectrum near 3.2 eV owing to the energetic isolation of O 2$p$ and transition metal $d$ bands. The strong electron correlation in these materials flattens the conduction band and increases $m^*$, thereby reducing the energy scale of intraband $d$–$d$ transitions (lower-lying $t_{2g}$→ higher-lying $e_g$ states and $t_{2g}$→$t_{2g}$ in $d$ bands under an octahedral crystal field) and simultaneously limiting absorption on the red end of the visible spectrum near 1.75 eV. Consequently, a transparent window in the visible light region is achieved. Furthermore, owing to the high $m^*$, a large $n$ of ~ $10^{23}$ cm$^{-3}$ can be allowed while maintaining the screened plasma energy below the visible range (<1.75 eV), which is in contrast to the case of conventional metals. Although the high $m^*$ negatively influences the conductivity relative to metals, the conductivity is better than that of degenerately doped wide-bandgap semiconductors [7]. This "correlated metal" approach potentially provides a better trade-off between the optical transparency and electrical conductivity while also alleviating the dependency on highly priced indium [1].

Among the TCOs with correlated metallic ground states, SrVO$_3$ (SVO) has been studied extensively and is expected to have practical applications owing to its high performance [7–11]. However, owing to the high carrier concentration of SVO, weak but significant plasma absorption is observed around the lower limit of the visible light range (~1.75 eV) [7,8,11]. In addition, there is noticeable optical absorption in the higher-energy region of the visible spectrum starting from 2.8 eV [7,8,11]. Therefore, more suitable materials for TCOs are desired, especially those with controllability of $n/m^*$ and larger interband transition energy to reduce absorption in the lower- and higher-energy regions of the visible spectrum, respectively.

To arbitrarily control $n/m^*$ as well as the energy range of interband transition in the higher-energy region, the use of SrTi$_{1-x}$V$_x$O$_3$ (STVO), which is a solid solution of SVO and SrTiO$_3$ (STO), has been proposed as a possible candidate [16]. STO with $3d^0$ configuration is a typical oxide semiconductor with a wide gap of 3.2 eV [17–20]. Although STO is fundamentally a TCO in the traditional semiconductor approach, its performance does not meet the requirements of general applications owing to the small overlap between Ti $3d$ bands and the resultant low mobility at room temperature [21]. On the contrary, SVO is an archetypal Fermi-liquid metal with simple $3d^1$ configuration [22–25] and has been extensively studied as one of the most promising TCO candidates for the correlated metal approach described above [7–11,26–28]. From the band structure perspective, SVO can be described as "extremely" doped STO, i.e., the fundamental band structures are similar, but the Fermi level ($E_F$) position is higher (approximately 0.5 eV) in SVO than STO owing to the accommodation of one $3d$ electron per unit cell in the conduction band [19,20,29–31]. Therefore, a solid solution of STO and SVO will have the capability of controlling the band filling by varying $x$. In other words, $n/m^*$ can be adjusted to an appropriate value by finetuning of $x$, assuming that the change in $m^*$ is less than that in $n$ under the framework of rigid-band picture. Furthermore, recent theoretical work has predicted the tunability of the interband transition in the higher-energy region of STVO [16], which reduces absorption in the higher-energy region of the visible spectrum, starting from 2.8 eV in SVO. In this study, to test the feasibility of STVO for finetuning the performance of transparent conduction, we investigated the electronic structure of STVO via *in situ* photoemission spectroscopy (PES) measurements, where $n$ and $m^*$ were experimentally determined. We will discuss the change in electronic structure of STVO with the substitution of V with Ti and the feasibility of STVO for TCO applications in terms of the observed electronic structures.

## II. EXPERIMENTAL

Epitaxial STVO thin films with a thickness of approximately 40 nm were grown on the (001) surface of 0.05 wt.% Nb-doped STO substrates in a laser molecular-beam epitaxy (MBE) chamber connected to an *in situ* photoemission system at BL-2A MUSASHI of the Photon Factory, KEK [32]. Sintered SrTi$_{1-x}$V$_x$O$_y$ ($x$ = 1.0, 0.8, 0.6, and 0.4) pellets were used as ablation targets. An Nd-doped yttrium aluminum garnet laser was used for target ablation in its frequency-tripled mode ($\lambda$ = 355 nm) at a repetition rate of 1 Hz. During the deposition, the substrate temperature was maintained at 900°C, and the oxygen pressure was maintained at less than $10^{-8}$ Torr. During film growth, the intensity of the specular spot in the reflection high-energy electron diffraction (RHEED) pattern was monitored to determine the film growth rate. The layer-by-layer growth of the STVO films was confirmed by the observation of clear RHEED oscillations. After cooling to below 100°C, the films were moved into the photoemission chamber under an ultrahigh vacuum of $10^{-11}$ Torr. In-vacuum transfer was necessary to maintain a high-quality surface.

All spectroscopic measurements were conducted *in situ* at 20 K. PES measurements in the soft-x-ray region were performed using a VG-Scienta SES-2002 analyzer with total energy resolutions of 150 and 300 meV at photon energies ($h\nu$) of 600 and 1200 eV, respectively. X-ray absorption spectroscopy (XAS) was also performed in total electron yield mode. *In situ* angle-resolved photoemission spectroscopy (ARPES) experiments were conducted using two orthogonally linear polarizations from incident light with $h\nu$ = 88 eV in the vacuum ultraviolet (VUV) light region, as shown in Fig. S1 in the Supplemental Material [33]. In these experiments, the energy and angular resolutions were set to approximately 30 meV and 0.3°, respectively. The $E_F$ of the samples was inferred from gold foil in electrical contact with the sample. The surface structure and cleanness of the vacuum-transferred STVO films were examined by low-energy electron diffraction (LEED) and core-level photoemission measurements. No detectable C 1$s$ peak was observed in the core-level photoemission spectra. These results indicate that no cleaning procedure was required for the *in situ* spectroscopic measurements. The stoichiometry of the samples was carefully characterized by analyzing the relative intensity of the relevant core levels, confirming that the cation composition of the samples was the same as that of the targets. The results of the detailed characterization of the grown STVO films are presented in the Supplemental Material [33].

The surface morphology of the measured STVO thin films was analyzed by *ex situ* atomic force microscopy in air, and atomically flat step-and-terrace structures were observed (see Fig. S2 in the Supplemental Material [33]). The crystal structure was characterized by x-ray diffraction, which confirmed the coherent growth of single-phase STVO films on the substrates (see Fig. S3 in the Supplemental Material [33]). The electrical resistivity was measured using the standard four-probe method. The samples for the resistivity measurements were grown on $(LaAlO_3)_{0.3}-(SrAl_{0.5}Ta_{0.5}O_3)_{0.7}$ (LSAT) substrates to prevent electric current from flowing through the conducting substrate (see Fig. S4 in the Supplemental Material [33]). The transport properties are in good agreement with those previously reported for STVO films on LSAT substrates [36].

### III. RESULTS
#### A. Valence and conduction band structures

Figure 1 shows the valence band and O-1$s$ XAS spectra for the STVO films with different $x$ values grown on Nb:STO substrates, representing the electronic states below and above $E_F$. These spectra exhibit remarkable and systematic changes as a function of $x$. The valence band mainly consists of three structures: two prominent O 2$p$-derived structures exist at binding energies of 3.0–9.0 eV, while a characteristic structure emerges near $E_F$ of the STVO films [37–39]. The structure near $E_F$ is attributable to the V 3$d$ states based on the V 2$p$-3$d$ resonant photoemission measurements (not shown). As can be seen in Fig. 1(a), the V 3$d$ states are energetically well-isolated from the O 2$p$ band, leading to interband transition starting from the blue region of the visible spectrum [7,8,11]. Focusing on the O 2$p$ states, the leading edge gradually shifts to a higher binding energy with decreasing $x$, suggesting an increase in the energies of the interband optical transitions (O 2$p$ → transition-metal $d$ band), namely the reduction of absorption in the blue region of the visible spectrum [16]. Meanwhile, the valence band spectrum of Nb:STO substrates, corresponding to STVO $x = 0$, exhibits a band gap of 3.2 eV below $E_F$, reflecting the *n*-type semiconducting nature of Nb:STO [19]. In the case of SVO ($x = 1.0$) films, there are two components attributable to the V 3$d$ states: a peak located precisely at $E_F$ and a relatively broad peak centered at approximately 1.5 eV, which correspond to the coherent (quasiparticle peak) and incoherent (remnant of the lower Hubbard band) parts, respectively [37–39]. The existence of the two-peak structure in the STVO films reflects the strong electron correlation in the STVO films. The relative intensity of the two-peak structure seems to change significantly with changing $x$, as discussed in detail later. Note that the spectral change near $E_F$ is not described by linear combination of the SVO and STO spectra. This indicates that a solid solution is achieved in the investigated STVO films across the composition range of $0.4 \leq x \leq 1$, which is consistent with the fact that the change in lattice constants of the STVO films follows Vegard's law [40] (see Fig. S3 in the Supplemental Material [33]).

For the conduction band, the O-1$s$ XAS spectra shown in Fig. 1(b) [41,42] also exhibit systematic changes as a function of $x$. According to previous works on SVO [39] and STO [43], the sharp peak around $E_F$ and broad peak at approximately -2.4 eV are attributable to unoccupied V 3$d$ $t_{2g}$ and $e_g$ states for STVO films, while the broad peaks at approximately -1.9 eV and -4.3 eV are attributable to Ti 3$d$ $t_{2g}$ and $e_g$ states, respectively. With decreasing $x$, the intensity of the V 3$d$ states decreases, while that of the Ti 3$d$ states increases. The energy positions of the V 3$d$ and Ti 3$d$ states remain almost unchanged for STVO films in the composition range of $x$ = 1.0–0.4, reflecting the pinning of $E_F$ within the V 3$d$ conduction band in STVO. These results indicate that the V 3$d$ $t_{2g}$ states mainly contribute to the conduction of the STVO films. In contrast, the energy position of the Ti 3$d$ states of Nb:STO deviates far from the chemical trend of STVO, reflecting its $n$-type semiconducting nature. These results suggest the existence of some kind of metal-to-semiconductor crossover in the conducting mechanism of STVO when $x$ ranges from 0.4 to 0.

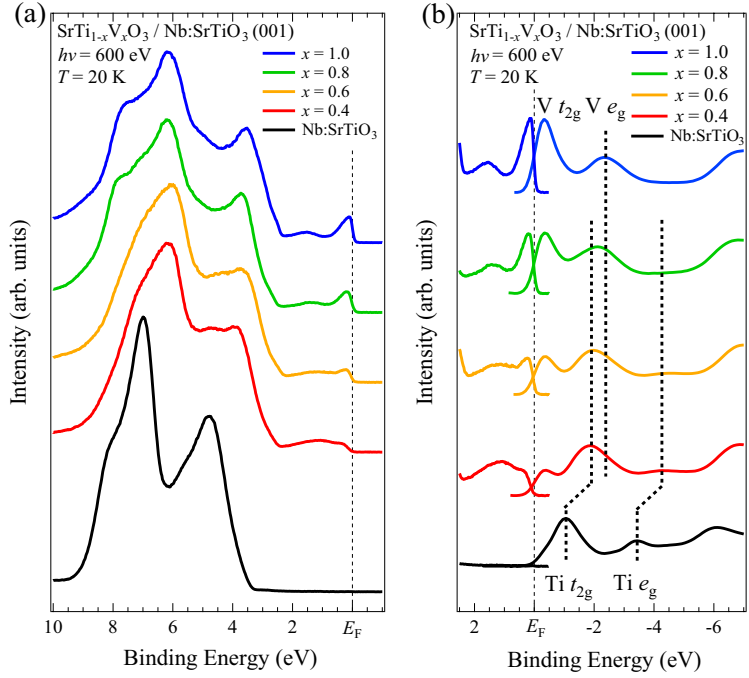

**FIG. 1.** (a) *In situ* valence band spectra of STVO thin films with varying $x$ grown on Nb:STO substrates, alongside those of Nb:STO substrate for reference. (b) Combined plot of PES spectra near $E_F$ and O-1$s$ XAS spectra of STVO thin films. It is well known that the $E_F$ position for the O-1$s$ XAS spectra cannot be determined unambiguously from O 1$s$ core-level PES and XAS data because of the unknown effects of the core-hole potential [41,42]. Thus, these XAS spectra have been aligned as follows. The O-1$s$ XAS spectrum of SVO is aligned to reflect its metallic nature [37–39], and the $E_F$ positions of the XAS spectra of STVO with various $x$ were determined by combining the $E_F$ position in the SVO spectrum with the $x$-dependent shift of the O 1$s$ core-level peak for the sake of convenience. For Nb:STO, the XAS spectrum has been aligned to reflect its energy gap of 3.2 eV.

To investigate the changes occurring in the V 3$d$ states near $E_F$ in more detail, we recorded the PES spectra near $E_F$, as shown in Fig. 2(a). For SVO films ($x = 1.0$), the prominent peak located just at $E_F$ indicates its metallic nature, which is consistent with a previous report [37]. According to the substitution of V with Ti (with decreasing $x$), the intensity of the coherent peak decreases by transferring its spectral weight to the incoherent states around 1.0–1.5 eV, indicating an increase in electron–electron correlation. Simultaneously, the leading edge of the V 3$d$ states appears to shift from above to below $E_F$ at $x = 0.8$, suggesting the evolution of a pseudogap (suppression of spectral weight) at $E_F$. With a further decrease in $x$, the spectral weight at $E_F$ becomes negligible, and an energy gap seems to open at $x = 0.4$. Although there is a negligibly small residual spectral weight at $E_F$ for the $x = 0.4$ film, an extrapolation of the linear portion of the leading edge to the energy axis yields a valence-band maximum of approximately 100 meV for the STVO film, indicating its insulating nature. The occurrence of the composition-driven metal–insulator transition (MIT) is further confirmed by the plot of the spectral weight at $E_F$ in the energy window of ± 10 meV as a function of $x$: the spectral weight at $E_F$ steeply decreases from $x = 1.0$ to 0.4 and approaches close to 0 at $x = 0.4$. Thus, these results indicate the occurrence of MIT at $x = 0.4$–0.6.

Although an energy gap seems to open at $x = 0.4$ owing to the depletion of spectral weight at $E_F$, the coherent peak structure still remains at approximately 0.3 eV. This feature seems to be different from that of the typical MIT of strongly correlated oxides [44,45]. In the case of MIT due to strong electron correlation, the Mott–Hubbard gap with the energy scale of the Coulomb interaction ($U \sim$ a few eV) opens at $E_F$ by spectral weight transfer from the coherent peak at $E_F$ to the Hubbard bands [44–47]. The energy scale of the observed gap is much smaller than that of Mott insulators. In addition, for the spectral weight transfer observed in strongly correlated oxides, the metallic edge at $E_F$ itself remains in the metallic region [44,45]. Thus, the observed spectral behavior suggests that the MIT of STVO is not explained by the strong electron correlation effect alone.

Liu *et al.* reported the electronic structure of STVO solid solutions based on *ab initio* density functional theory (DFT) with generalized gradient approximation (GGA) + $U$ correction [16]. They found that the substitution of V with Ti ions caused composition-driven MIT in STVO at close to $x = 0.67$. According to the calculation, the density of states (DOS) of STVO in the energy range of 10–-5 eV mainly consists of V 3$d$, Ti 3$d$, and O 2$p$ states. The V 3$d$ $t_{2g}$ and $e_g$ states are located in the near-$E_F$ region (binding energy of 1 eV to -1.5 eV) and the region from -1.0 to -3.5 eV, respectively, while the Ti 3$d$ states extend above -2.0 eV and the O 2$p$ states mainly exist below 2 eV. Thus, the V 3$d$ $t_{2g}$ states mainly accommodate conduction electrons and dominate the conductivity of STVO films, which is consistent with our experimental results. With decreasing $x$, the partial DOS of the V 3$d$ $t_{2g}$ states in the vicinity of $E_F$ was reduced and the energy gap opened at $x = 0.67$ by the partial DOS transfer from $E_F$ to the lower and upper Hubbard bands. The observed composition-derived MIT is in good agreement with the prediction of band structure calculation, although there is a quantitative discrepancy in critical $x$ for composition-derived MIT. On the contrary, the pseudogap behavior at $E_F$ and the remnant coherent peak at 0.3 eV were not reproduced by this calculation [16].

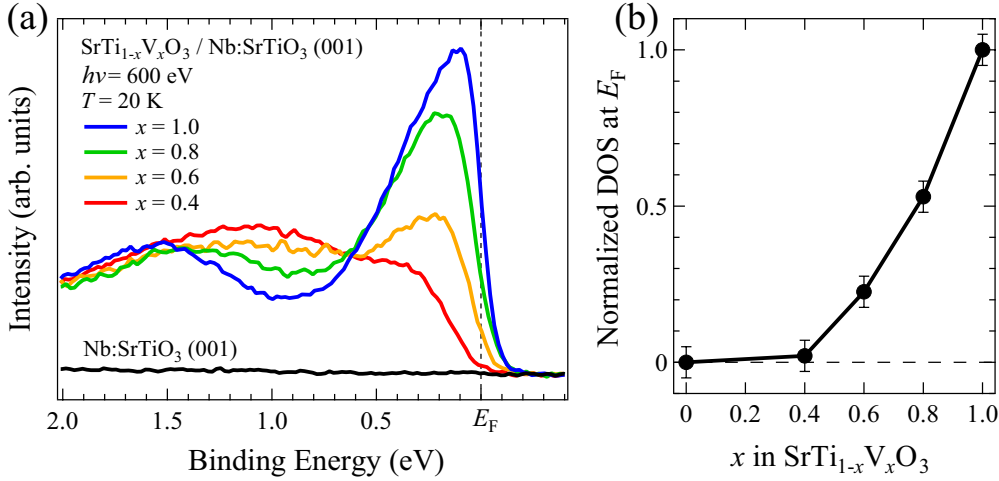

**FIG. 2.** (a) *In situ* PES spectra near $E_F$ of STVO thin films with different $x$ grown on Nb:SrTiO$_3$ substrates, alongside those of Nb:SrTiO$_3$ substrate for reference. (b) Plot of DOS at $E_F$ as a function of $x$. Note that for $x \leq 0.4$, DOS at $E_F$ would be almost 0 owing to the MIT at $x = 0.4$–0.6.

## B. V 3d band structure near $E_F$

Although the overall electronic structure of STVO is in good agreement with the prediction of band structure calculations, there are significant discrepancies in the spectral changes near $E_F$. To obtain better insight into the spectral behavior near $E_F$, we performed *in situ* ARPES measurements of the STVO films. Figure 3 shows the ARPES intensity map for STVO films for $x = 0.4$–1.0 taken along the Γ–X direction (see Fig. S5 in the Supplemental Material [33]) together with the momentum distribution curve (MDC) at $E_F$ and energy distribution curve (EDC) at the Γ point. According to the band structure calculations [48], the Fermi surface (FS) sheets of bulk SVO with cubic symmetry are essentially formed from three intersecting cylinders containing the V 3d $d_{xy}$, $d_{yz}$, and $d_{zx}$ states. Owing to the characteristic FS topology of SVO, $m^*$ at the Fermi momentum ($k_F$) in the direction of the diameter of the cylinder's circle, namely the Γ–X direction, roughly represents the overall value of $m^*$ in SVO (see Fig. S5 in the Supplemental Material [33]). The band structures of SVO along the Γ–X direction consist of three bands: two degenerate parabolic dispersions derived from the $d_{xy}$ and $d_{zx}$ states and a nearly nondispersive $d_{yz}$ state [23–25,29–31,49]. Note that the ARPES images shown in Fig. 3 consist of only the $d_{zx}$ bands of V 3d $t_{2g}$ states in the present experimental geometry owing to the dipole selection rules for the light polarizations and each orbital symmetry with respect to the mirror plane (see Fig. S1 in the Supplemental Material [33]).

As can be seen in Fig. 3, a clear parabolic band dispersion is observed for the metallic region of $x = 0.6$–$1.0$. As $x$ decreases, the conduction band minimum (CBM) approaches $E_F$, and the $k_F$ consequently decreases. This systematic behavior demonstrates that the fundamental electronic structure of STVO in the metallic region is described by the framework of the rigid-band model; the CBM gradually shifts by 0.1 eV from $x = 1.0$ to 0.6 (see also the plot in Fig. 4(a)). In contrast, the ARPES intensity in the vicinity of $E_F$ significantly weakens with approaching the insulating region and almost disappears at $x = 0.4$, although a faint parabolic band remains in the ARPES images. The steep suppression in the quasiparticle intensity is highlighted in Fig. 4(b), where the ARPES spectra at $k_F$ are plotted (see also Fig. S6 in the Supplemental Material [33]). In addition, accompanying the decrease in spectral intensity, there is an increase in intensity on the high binding energy side, which is consistent with the behavior of the angle-integrated PES spectra shown in Fig. 2.

The conduction band in the vicinity of $E_F$ gradually smears out with decreasing $x$ and almost disappears in the insulating phase ($x = 0.4$), whereas the energy position of the conduction band monotonically shifts in a rigid-band manner, and the band dispersion itself remains unchanged in the metallic phase, as can be seen in Fig. 3. Such a gradual disappearance of the FS near MIT is commonly observed in strongly correlated oxides [50–53]; the FS gradually disappears near MIT by transferring spectral weight from the coherent band near $E_F$ to the higher binding-energy side, whereas the band dispersion itself remains unchanged. Thus, these results indicate that the spectral weight transfer dominates the pseudogap or gap formation in STVO associated with the composition-driven MIT. These spectral behaviors are not described within the framework of the one-electron picture, although the overall electronic structure of STVO is in good agreement with the prediction of band structure calculations.

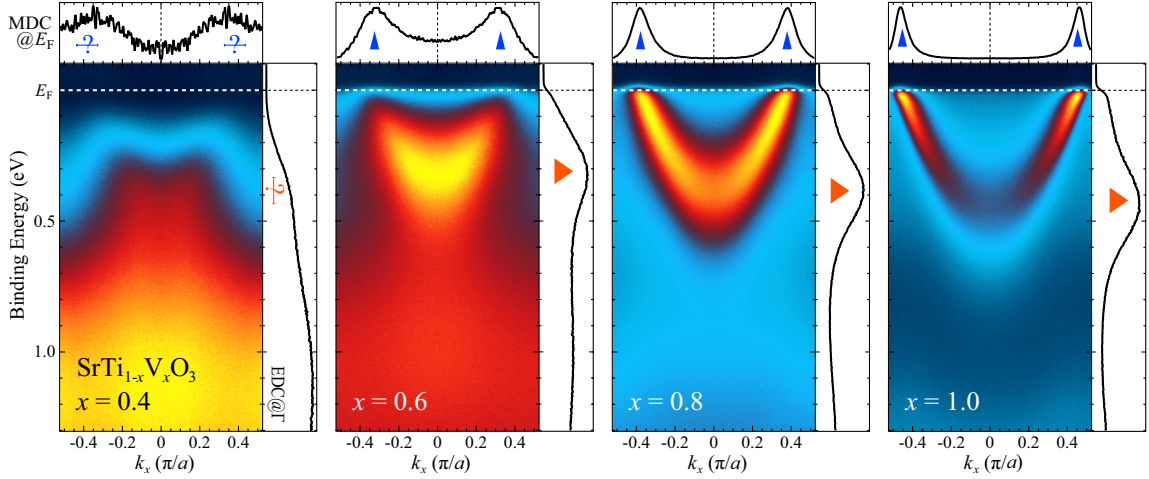

**FIG. 3.** *In situ* ARPES intensity plot along the $k_x$ direction for STVO with $x = 0.4$–$1.0$. The intensity plots are symmetrized with respect to the Γ point ($k_x = 0$) and averaged [33]. The MDCs at $E_F$ with an energy window of ±10 meV are shown in the respective ARPES images (top panel), while the EDCs at the Γ point (along the center line ($k_x = 0$)) are on the right side of each image. The orange arrows indicate the bottom of the conduction band, while blue ones in MDCs indicate $k_F$. Question marks represent considerably weak structures. Note that the ARPES images consist of only the $d_{zx}$ bands of V $3d$ $t_{2g}$ states in the present experimental geometry owing to the dipole selection rules for the light polarizations and each orbital symmetry with respect to the mirror plane [33].

## C. Estimation of parameters for transparent conduction

To address the key factor of $n/m^*$ in optimizing the performance of transparent conducting oxides, we evaluated each parameter of $n$ and $m^*$ from the ARPES results, as shown in Fig. 4. First, we estimated $n$ of the STVO films from $k_F$ obtained from the ARPES measurements, assuming three simple ($d_{xy}$, $d_{yz}$, and $d_{zx}$) cylindrical FSs intersecting at the Γ points [48]. The results are plotted in Fig. 4(c) together with $k_F$ as a function of $x$. The value of $n$ seems to be proportional to $x$ in the range of 1.0 to 0.6, suggesting that the change in the V $3d$ states is essentially rigid-band like in this composition range. As a result, $n$ is reduced by 52±10% at $x = 0.6$, as expected, although it seems to increase again at $x = 0.4$. The strange spectral behavior at $x = 0.4$ is discussed later.

Second, from the obtained band dispersion, we also evaluated the change in the effective mass of the STVO films. The band renormalization factor $Z$, whose inverse ($1/Z$) corresponds to the mass enhancement factor, was estimated by fitting the obtained dispersion using the following procedure.

We employed the simplified $t_{2g}$ tight-binding (TB) band dispersion for the corresponding $d_{zx}$ band of $E_{zx}(k_x)$:

$$E_{zx}^{TB}(k_x) = \varepsilon_{t_{2g}} + 2t_\pi(\cos k_x + 1) + 2t_\delta + 4t'_\sigma \cos k_x$$

Eq. (1)

where $\varepsilon_{t_{2g}}$ is the energy of the three degenerate $t_{2g}$ atomic (Wannier) orbitals. $t_\pi$ and $t_\delta$ are the first-nearest neighbor hopping parameters, and $t'_\sigma$ is the second-nearest neighbor hopping parameter. The tight-binding parameters of $\varepsilon_{t_{2g}}$= 625 meV (1,474 meV), $t_\pi$ = -281 meV (-277 meV), $t_\delta$ = -33 meV (-31 meV), and $t'_\sigma$ = -96 meV (-76 meV) were determined by fitting the results of the band structure calculation based on the local-density approximation for SVO (STO) [48]. For the TB parameters for solid solution STVO films ($x$ = 0.8, 0.6, and 0.4), we used values estimated from the interpolation of the two parent material values because the lattice constant of STVO films obeys Vegard's law [33]. Then, we fitted the experimental band dispersion using the following equation:

$$E(k_x) = Z \cdot E_{zx}^{TB}(k_x) + \varepsilon_0$$

Eq. (2)

where $\varepsilon_0$ is an adjustment parameter [49].

Figure 5 shows a comparison of the band structure of STVO films obtained from the peak positions of the EDCs and MDCs with the TB results. As shown in Fig. 5, the fitted TB curve is in good agreement with the experimental data, validating our analysis for determining the band renormalization. The estimated 1/$Z$ is plotted in Fig. 4(d) as a function of $x$. As expected from the rigid-band-like behavior of the V 3$d$ band in STVO (Fig. 3(a)), the change in the mass enhancement factor shows a fairly weak $x$-dependence, in contrast to $n$. The essential rigid-band behavior of the V 3$d$ states is further supported by the plot of the values for CBM and $n$ vs. $x$, as shown in Figs. 4(a) and 4(c), respectively. These results suggest that the key factor of $n/m^*$ for optimizing the performance of transparent conducting materials is tuned by $x$.

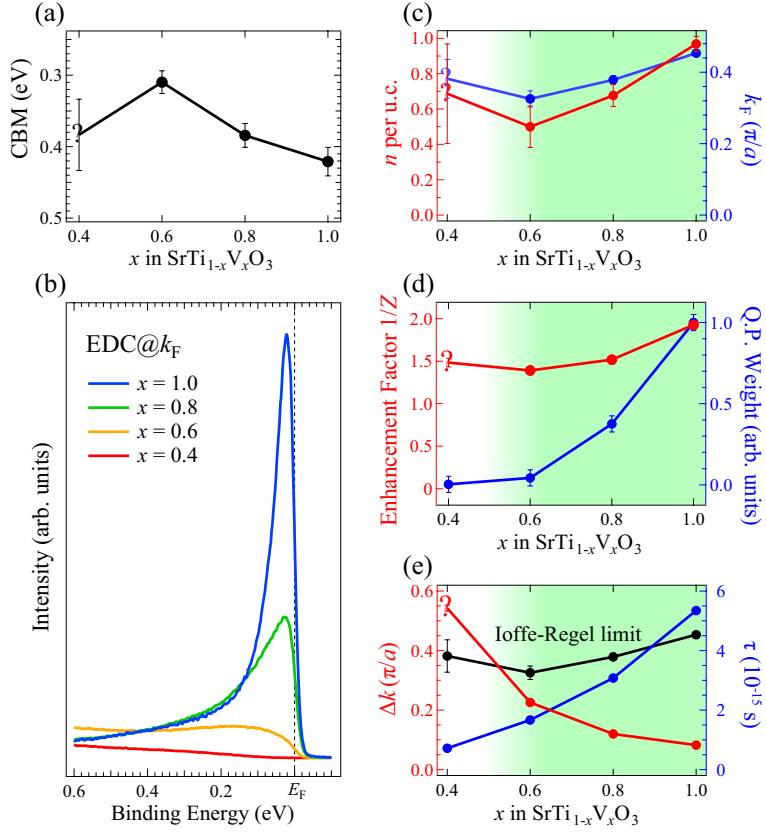

**FIG. 4** Plots of physical parameters obtained from ARPES measurements: (a) The plot of CBM of V 3$d$ $t_{2g}$ band as a function of $x$. (b) ARPES spectra (EDCs) at $k_F$. The plot of (c) $n$ and $k_F$, (d) $1/Z$ and quasiparticle (QP) spectral weight at $E_F$, and (e) $\Delta k$ and $\tau$ as a function of $x$. For (e), the Ioffe-Regel (IR) limit of corresponding states is also shown. The gradation area is the composition-derived MIT. Note that the values of $k_F$ and $\Delta k$ were determined by fitting the MDCs shown in the top panel of Fig. 3 to the linear combination of Lorentzians and a smooth background.

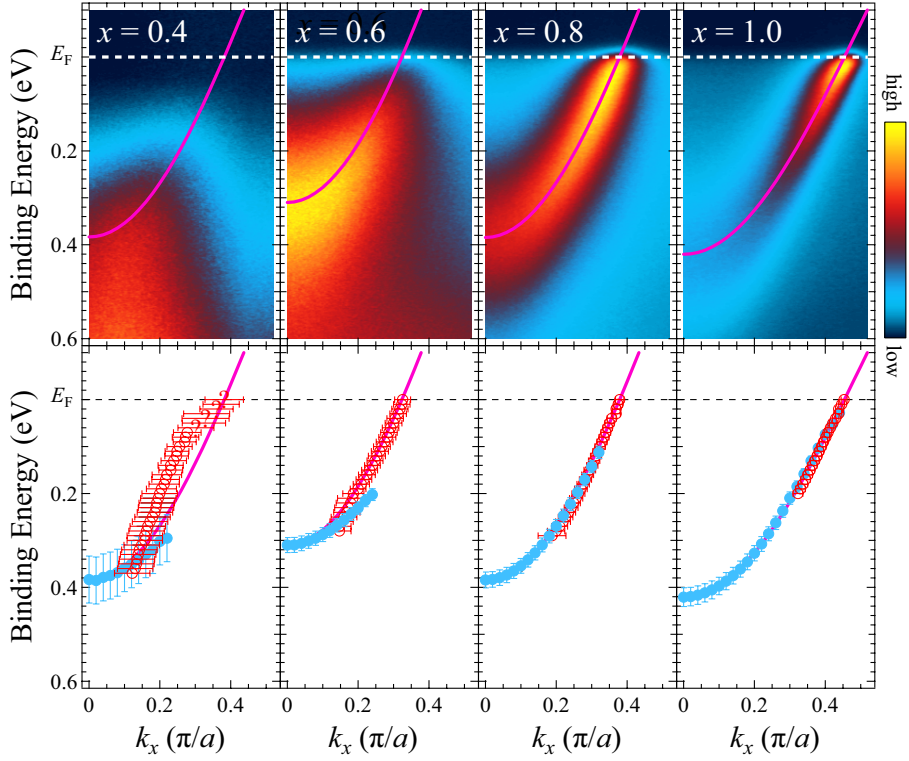

**FIG. 5**. (Upper panel) Comparison of ARPES intensity plot for STVO films with the corresponding results of TB fitting by Eq. (2). (Lower panel) Peak positions determined based on EDCs and MDCs shown by open red and filled blue circles, respectively. Question marks represent considerably weak structures. The results of the TB fitting are also overlaid.

## IV. DISCUSSION

### A. Origin of the composition-driven MIT in STVO

Here, we discuss the origin of the composition-driven MIT in STVO. As can be seen in Figs. 3 and 4, the spectral behavior of the insulating region of $x = 0.4$ deviates from the rigid-band behavior in the metallic region of $x = 0.6$–$1.0$. Meanwhile, faint parabolic band dispersion is observed at $x = 0.4$, where the ARPES intensity weakens approaching $E_F$ and nearly disappears at $E_F$. Consequently, faint peaks seem to exist in the MDCs. The remnant peaks in the MDCs are indicative of the FS of STVO $x = 0.4$, even though it is in insulating states.

Considering that $1/Z$ is almost unchanged with $x$, the suppression of the quasiparticle peaks may not only be due to the strong electron–electron correlation because the quasiparticle weight at $k_F$ should be scaled with $1/Z$ in the case that only the strong electron–electron correlation is a matter [45–47,54,55]. As can be seen in Figs. 4(b) and 4(d), the quasiparticle peak at $E_F$ significantly reduces its intensity, while the width of the MDC peak at $E_F$ [$\Delta k(E_F)$] simultaneously becomes relatively broader with decreasing $x$, as summarized in the plot of Fig. 4(e). These spectral behaviors imply a different origin of the composition-derived MIT in STVO from the Mott–Hubbard gap. Because $\Delta k$ reflects the mean free path $l$ of conduction electrons as $l = v_F \cdot \tau = 1/\Delta k$, where $v_F$ and $\tau$ are the Fermi velocity and scattering time, the broadening of $\Delta k$ suggests a smaller $\tau$ [56]. Thus, the significant broadening of $\Delta k$ suggests the occurrence of a disorder-driven MIT in STVO [34]. To test the possible disorder origin of the MIT, we overlaid the Ioffe–Regel (IR) limit of $k_F/\Delta k(E_F) \sim 1$ [57] in Fig. 4(e). In the metallic region, $\Delta k$ in SVO is well below the IR limit and approaches the IR limit as $x$ decreases towards the critical composition of the MIT. Eventually, the $\Delta k$ value is well above the IR limit for $x = 0.4$, indicating that the effects of the disorder are sufficiently strong to cause MIT in the studied STVO films. Therefore, based on the ARPES analysis, it can be naturally concluded that the composition-driven MIT in STVO is dominantly governed by the strong disorder due to chemical substitution.

Although the origin of the strange behavior at $x = 0.4$ is not clear at the moment, a plausible interpretation is that it is due to the interplay between strong electron–electron interactions and disorder-induced localization. In STVO with $Ti^{4+}$ ($3d^0$) and $V^{4+}$ ($3d^1$) ions randomly occupying the transition metal sites, the $Ti^{4+}$ ions perturb the periodic potential of the V $3d$ band, which introduces significant disorder and Anderson-localized states [34,58–60]. Meanwhile, STVO is a strongly correlated electron system with $1/Z = 1.4–1.9$. The observed spectral behavior and related composition-dependent MIT may be understood in terms of the combined effects of electron correlations and disorder potentials in STVO. Indeed, similar pseudogap formation has been observed in coherent states for disordered strongly correlated oxides [61–64]. However, to properly elucidate the spectral behavior associated with both effects, further investigation is necessary.

**B. Screening as a strongly correlated transparent electrode**
Finally, we briefly discuss the feasibility of precise optimization of the transparent conductive performance of STVO films. From the detailed analysis of the photoemission spectra, we revealed that 1) the overall electronic structure of STVO is described in the framework of the band structure calculation, while the spectral behavior for V 3$d$ states near $E_F$ is not; 2) the spectral weight of the coherent band near $E_F$ is suppressed by spectral weight transfer; 3) simultaneously, the pseudogap is evolved at $E_F$; 4) as a result of spectral weight transfer and depletion of spectral weight at $E_F$, composition-driven MIT occurs at $x$ = 0.4–0.6; 5) in metallic STVO ($x$ = 1.0 to 0.6), carrier concentration $n$ in the metallic region is proportionally reduced from 0.97 ($x$ = 1.0) to 0.50 ($x$ = 0.6); and 6) in contrast to $n$, the value of 1/$Z$, which corresponds to $m^*$, is slightly varied from 1.9 ($x$ = 1.0) to 1.4 ($x$ = 0.6). These results demonstrate the possibility of controlling the $n/m^*$ value by approximately 28 % in the composition range of $x$ = 1.0–0.6, namely the controllability of the plasma absorption edge to an appropriate value by tuning $x$. Controllability has important implications for the future development of strongly correlated oxide transparent electrodes. For the strongly correlated TCOs that have been studied so far [7–14], such as SVO and CVO, finetuning of the properties is not possible. We have demonstrated that finetuning of properties is possible in solid solution form by tuning an appropriate composition ratio in the same manner as TCOs based on wide-gap semiconductors. Meanwhile, high doping (substitution of V with Ti) with $x$ inherently induces randomness in the crystal, resulting in reduced mobility.

In contrast, by mixing SVO with STO with larger interband optical transitions (O 2$p$ → transition metal $d$ band) of 3.2 eV, the reduction of absorption in the blue spectra may be reduced in STVO [7,8]. In fact, the O 2$p$ states in the valence band spectra rigidly shift to the higher binding energy side by 300 meV (Fig. 1) with decreasing $x$ from 1.0 to 0.4, whereas the bottom of the V 3$d$ conduction band oppositely shifts by 100 meV (Fig. 4). These spectral behaviors suggest the energy increment of the interband optical transition in STVO, as predicted from the DFT calculation [16]. Therefore, the present spectroscopic screening of STVO as a transparent electrode material might be a promising direction for research on TCOs based on strongly correlated oxides as well as a fundamental understanding of strongly correlated oxides. However, we must consider the effects of complex electron–electron scattering in such strongly correlated oxides. To optimize the performance of strongly correlated TCO, we must pay attention to the delicate trade-off relation between the abovementioned merits from the substitution and the deterioration of the conductivity due to disorder. In addition, to gain a more comprehensive understanding of the electronic structure of solid solutions of strongly correlated oxides, further investigation is required. In particular, the effects of different constituent transition metals on the electronic structure and the resultant transparent performance are especially important to examine.

## V. CONCLUSION

We investigated the electronic structure of STVO thin films by *in situ* PES to test their potential for use as strongly correlated TCO with tunable performance. One of the possible approaches for controlling the performance is the use of such solid solutions between the strongly correlated TCO $SrVO_3$ and oxide semiconductor $SrTiO_3$. From the PES and O-1*s* XAS spectra, systematic and significant spectral changes were found near $E_F$, although the overall electronic structure of STVO is in good agreement with the prediction of band structure calculations. As $x$ decreases from 1.0, spectral weight transfer occurs from the coherent band near $E_F$ to the incoherent states (lower Hubbard band) around 1.0–1.5 eV in the PES spectra. Simultaneously, a pseudogap is formed at $E_F$, which corresponds to the significant reduction in the quasiparticle spectral weight of the ARPES spectra at the Fermi momentum. The pseudogap finally evolves into an energy gap at $x = 0.4$, indicating the occurrence of a composition-driven MIT.

Meanwhile, from the ARPES results, the carrier concentration $n$ changes proportionally as a function of $x$. In contrast, the mass enhancement factor, which is proportional to the effective mass ($m^*$), does not change significantly with varying $x$ in the metallic range of $x = 0.6$–1.0. These results suggest that the electronic structure of STVO in the metallic region is described by the framework of the rigid-band model, and consequently, the key factor of $n/m^*$ in optimizing the performance of transparent conducting materials is tuned by $x$. However, significant broadening of the MDC peak width at $E_F$ ($\Delta k$) is observed with decreasing $x$, reflecting the increase in the scattering rate. The detailed analysis reveals that $\Delta k$ at $x = 0.4$ is well above the IR limit, demonstrating that the composition-driven MIT in STVO films predominantly originates from disorder-induced Anderson localization due to chemical substitution. These results suggest that the control of TCO performance in strongly correlated metals using solid solutions is effective in a certain region near the end composition. However, care must be taken in adopting this for a wider range of compositions because the unavoidable chemical disorder effects in strongly correlated materials also drive the MIT.

## ACKNOWLEDGMENTS

The authors are very grateful to Y. Kuramoto and M. Kobayashi for their helpful discussions and acknowledge M. Minohara for his advice on sample growth. This work was financially supported by a Grant-in-Aid for Scientific Research (Nos. 16H02115, 16KK0107, and 20KK0117) from the Japan Society for the Promotion of Science (JSPS), CREST (JPMJCR18T1) from the Japan Science and Technology Agency (JST), and the MEXT Element Strategy Initiative to Form Core Research Center (JPMXP0112101001). The preliminary sample characterization using hard x-ray photoemission at SPring-8 was conducted with approval from the Japan Synchrotron Radiation Research Institute (2019B1248). The work performed at KEK-PF was approved by the Program Advisory Committee (proposals 2019T004 and 2018S2-004) at the Institute of Materials Structure Science, KEK.

# REFERENCES

[1] K. Ellmer, Past achievements and future challenges in the development of optically transparent electrodes, Nat. Photonics **6**, 809 (2012).

[2] H. Hosono, Recent progress in transparent oxide semiconductors: Materials and device application, Thin Solid Films **515**, 6000 (2007).

[3] K. Nomura, H. Ohta, A. Takagi, T. Kamiya, M. Hirano, and H. Hosono, Room-temperature fabrication of transparent flexible thin-film transistors using amorphous oxide semiconductors, Nature (London) **432**, 488 (2004).

[4] J. Shi, J. Zhang, L. Yang, M. Qu, D.-C. Qi, and K. H. L. Zhang, Wide Bandgap Oxide Semiconductors: from Materials Physics to Optoelectronic Devices, Adv. Mater., e2006230 33797084 (2021).

[5] A. J. Freeman, K. R. Poeppelmeier, T. O. Mason, R. P. H. Chang, and T. J. Marks, Chemical and Thin-Film Strategies for New Transparent Conducting Oxides, MRS Bull. **25**, 45 (2000).

[6] H. Mizoguchi, T. Kamiya, S. Matsuishi, and H. Hosono, A germanate transparent conductive oxide, Nat. Commun. **2**, 470 (2011).

[7] L. Zhang, Y. Zhou, L. Guo, W. Zhao, A. Barnes, H.-T. Zhang, C. Eaton, Y. Zheng, M. Brahlek, and H. F. Haneef, N. J. Podraza, M. H. Chan, V. Gopalan, K. M. Rabe, and R. Engel-Herbert, Correlated metals as transparent conductors, Nat. Mater. **15**, 204 (2016).

[8] M. Mirjolet, F. Sánchez, and J. Fontcuberta, High Carrier Mobility, Electrical Conductivity, and Optical Transmittance in Epitaxial $SrVO_3$ Thin Films, Adv. Funct. Mater. **29**, 1808432 (2019).

[9] J. L. Stoner, P. A. E. Murgatroyd, M. O'Sullivan, M. S. Dyer, T. D. Manning, J. B. Claridge, M. J. Rosseinsky, and J. Alaria, Chemical Control of Correlated Metals as Transparent Conductors, Adv. Funct. Mater. **29**, 1808609 (2019).

[10] A. Boileau, A. Cheikh, A. Fouchet, A. David, R. Escobar-Galindo, C. Labbé, P. Marie, F. Gourbilleau, and U. Lüders, Optical and electrical properties of the transparent conductor $SrVO_3$ without long-range crystalline order, Appl. Phys. Lett. **112**, 021905 (2018).

[11] M. Brahlek, L. Zhang, J. Lapano, H.-T. Zhang, R. Engel-Herbert, N. Shukla, S. Datta, H. Paik, and D. G. Schlom, Opportunities in vanadium-based strongly correlated electron systems, MRS Commun. **7**, 27 (2017).

[12] Y. Park, J. Roth, D. Oka, Y. Hirose, T. Hasegawa, A. Paul, A. Pogrebnyakov, V. Gopalan, T. Birol, and R. Engel-Herbert, $SrNbO_3$ as a transparent conductor in the visible and ultraviolet spectra, Commun. Phys. **3**, 102 (2020).

[13] L. Hu, R. Wei, J. Yan, D. Wang, X. Tang, X. Luo, W. Song, J. Dai, X. Zhu, C. Zhang, and Y. Sun, $La_{2/3}Sr_{1/3}VO_3$ Thin Films: A New p-Type Transparent Conducting Oxide with Very High Figure of Merit, Adv. Electron. Mater. **4**, 1700476 (2018).

[14] L. Hu, M. L. Zhao, S. Liang, D. P. Song, R. H. Wei, X. W. Tang, W. H. Song, J. M. Dai, G. He, C. J. Zhang, X. B. Zhu, and Y. P. Sun, Exploring High-Performance *p*-Type Transparent Conducting Oxides Based on Electron Correlation in $V_2O_3$ Thin Films, Phys. Rev. Appl. **12**, 044035 (2019).

[15] R. Wei, L. Zhang, L. Hu, X. Tang, J. Yang, J. Dai, W. Song, X. Zhu, and Y. Sun, *p*-type transparent conductivity in high temperature superconducting Bi-2212 thin films, Appl. Phys. Lett. **112**, 251109 (2018).

[16] Z. T. Y. Liu, N. J. Podraza, S. V. Khare, and P. Sarin, Transparency enhancement for $SrVO_3$ by $SrTiO_3$ mixing: A first-principles study, Comput. Mater. Sci. **144**, 139 (2018).

[17] D. Goldschmidt and H. L. Tuller, Fundamental absorption edge of $SrTiO_3$ at high temperatures, Phys. Rev. B **35**, 4360 (1987).

[18] K. van Benthem, C. Elsässer, and R. H. French, Bulk electronic structure of $SrTiO_3$: Experiment and theory, J. Appl. Phys. **90**, 6156 (2001).

[19] M. Takizawa, K. Maekawa, H. Wadati, T. Yoshida, A. Fujimori, H. Kumigashira, and M. Oshima, Angle-resolved photoemission study of Nb-doped $SrTiO_3$, Phys. Rev. B **79**, 113103 (2009).

[20] A. F. Santander-Syro, O. Copie, T. Kondo, F. Fortuna, S. Pailhès, R. Weht, X. G. Qiu, F. Bertran, A. Nicolaou, A. Taleb-Ibrahimi, P. Le Fèvre, G. Herranz, M. Bibes, N. Reyren, Y. Apertet, P. Lecoeur, A. Barthélémy, and M. J. Rozenberg, Two-dimensional electron gas with universal subbands at the surface of $SrTiO_3$, Nature (London) **469**, 189 (2011).

[21] J. Son, P. Moetakef, B. Jalan, O. Bierwagen, N. J. Wright, R. Engel-Herbert, and S. Stemmer, Epitaxial $SrTiO_3$ films with electron mobilities exceeding 30,000 $cm^2$ $V^{-1}$ $s^{-1}$, Nat. Mater. **9**, 482 (2010).

[22] I. H. Inoue, O. Goto, H. Makino, N. E. Hussey, and M. Ishikawa, Bandwidth control in a perovskite-type $3d^1$-correlated metal $Ca_{1-x}Sr_xVO_3$. I. Evolution of the electronic properties and effective mass, Phys. Rev. B **58**, 4372 (1998).

[23] S. Aizaki, T. Yoshida, K. Yoshimatsu, M. Takizawa, M. Minohara, S. Ideta, A. Fujimori, K. Gupta, P. Mahadevan, K. Horiba, H. Kumigashira, and M. Oshima, Self-Energy on the Low- to High-Energy Electronic Structure of Correlated Metal $SrVO_3$, Phys. Rev. Lett. **109**, 056401 (2012).

[24] M. Kobayashi, K. Yoshimatsu, E. Sakai, M. Kitamura, K. Horiba, A. Fujimori, and H. Kumigashira, Origin of the Anomalous Mass Renormalization in Metallic Quantum Well States of Strongly Correlated Oxide $SrVO_3$, Phys. Rev. Lett. **115**, 076801 (2015).

[25] M. Kobayashi, K. Yoshimatsu, T. Mitsuhashi, M. Kitamura, E. Sakai, R. Yukawa, M. Minohara, A. Fujimori, K. Horiba, and H. Kumigashira, Emergence of Quantum Critical Behavior in Metallic Quantum-Well States of Strongly Correlated Oxides, Sci. Rep. **7**, 16621 (2017).

[26] J. A. Moyer, C. Eaton, and R. Engel-Herbert, Highly Conductive $SrVO_3$ as a Bottom Electrode for Functional Perovskite Oxides, Adv. Mater. **25**, 3578 (2013).

[27] M. Brahlek, L. Zhang, C. Eaton, H.-T. Zhang, and R. Engel-Herbert, Accessing a growth window for SrVO$_3$ thin films, Appl. Phys. Lett. **107**, 143108 (2015).

[28] J. Wang, G. Rijnders, and G. Koster, Complex plume stoichiometry during pulsed laser deposition of SrVO$_3$ at low oxygen pressures, Appl. Phys. Lett. **113**, 223103 (2018).

[29] T. Yoshida, K. Tanaka, H. Yagi, A. Ino, H. Eisaki, A. Fujimori, and Z.-X. Shen, Direct Observation of the Mass Renormalization in SrVO$_3$ by Angle Resolved Photoemission Spectroscopy, Phys. Rev. Lett. **95**, 146404 (2005).

[30] T. Yoshida, M. Hashimoto, T. Takizawa, A. Fujimori, M. Kubota, K. Ono, and H. Eisaki, Mass renormalization in the bandwidth-controlled Mott-Hubbard systems SrVO$_3$ and CaVO$_3$ studied by angle-resolved photoemission spectroscopy, Phys. Rev. B **82**, 085119 (2010).

[31] M. Takizawa, M. Minohara, H. Kumigashira, D. Toyota, M. Oshima, H. Wadati, T. Yoshida, A. Fujimori, M. Lippmaa, M. Kawasaki, H. Koinuma, G. Sordi, and M. Rozenberg, Coherent and incoherent $d$ band dispersions in SrVO$_3$, Phys. Rev. B **80**, 235104 (2009).

[32] K. Horiba, H. Ohguchi, H. Kumigashira, M. Oshima, K. Ono, N. Nakagawa, M. Lippmaa, M. Kawasaki, and H. Koinuma, A high-resolution synchrotron-radiation angle-resoled photoemission spectrometer with *in situ* oxide thin film growth capability, Rev. Sci. Instrum. **74**, 3406 (2003).

[33] See Supplemental Material at [URL will be inserted by publisher] for geometry in the present ARPES measurements, detailed characterizations of the measured samples, the FS of SrVO$_3$, and the analysis of ARPES images, which includes Refs [34,35].

[34] P. Dougier, J. C. C. Fan, and J. B. Goodenough, Etude des proprietes magnetiques, electriques et optiques des phases de structure perovskite SrVO$_{2.90}$ et SrVO$_3$, J. Solid State Chem. **14**, 247 (1975).

[35] A. Okazaki and M. Kawaminami, Lattice constant of strontium titanate at low temperatures, Mater. Res. Bull. **8**, 545 (1973).

[36] M. Gu, S. A. Wolf, and J. Lu, Metal-insulator transition in SrTi$_{1-x}$V$_x$O$_3$ thin films, Appl. Phys. Lett. **103**, 223110 (2013).

[37] K. Yoshimatsu, T. Okabe, H. Kumigashira, S. Okamoto, S. Aizaki, A. Fujimori, and M. Oshima, Dimensional-Crossover-Driven Metal-Insulator Transition in SrVO$_3$ Ultrathin Films, Phys. Rev. Lett. **104**, 147601 (2010).

[38] I. H. Inoue, I. Hase, Y. Aiura, A. Fujimori, Y. Haruyama, T. Maruyama, and Y. Nishihara, Systematic Development of the Spectral Function in the $3d^1$ Mott-Hubbard System Ca$_{1-x}$Sr$_x$VO$_3$, Phys. Rev. Lett. **74**, 2539 (1995).

[39] K. Morikawa, T. Mizokawa, K. Kobayashi, A. Fujimori, H. Eisaki, S. Uchida, F. Iga, and Y. Nishihara, Spectral weight transfer and mass renormalization in Mott-Hubbard systems SrVO$_3$ and CaVO$_3$: Influence of long-range Coulomb interaction, Phys. Rev. B **52**, 13711 (1995).

[40] A. R. Denton and N. W. Ashcroft, Vegard's law, Phys. Rev. A **43**, 3161 (1991).

[41] T. Saitoh, A. E. Bocquet, T. Mizokawa, H. Namatame, A. Fujimori, M. Abbate, Y. Takeda, and M. Takano, Electronic structure of La$_{1-x}$Sr$_x$MnO$_3$ studied by photoemission and x-ray-absorption spectroscopy, Phys. Rev. B **51**, 13942 (1995).

[42] H. Wadati, D. Kobayashi, H. Kumigashira, K. Okazaki, T. Mizokawa, A. Fujimori, K. Horiba, M. Oshima, N. Hamada, M. Lippmaa, M. Kawasaki, and H. Koinuma, Hole-doping-induced changes in the electronic structure of La$_{1-x}$Sr$_x$FeO$_3$: Soft x-ray photoemission and absorption study of epitaxial thin films, Phys. Rev. B **71**, 035108 (2005).

[43] T. Higuchi, T. Tsukamoto, S. Yamaguchi, K. Kobayashi, N. Sata, M. Ishigame, and S. Shin, Observation of acceptor level of *p*-type SrTiO$_3$ by high-resolution soft-X-ray absorption spectroscopy, Nucl. Instrum. Methods Phys. Res. Sect. B-Beam Interact. Mater. Atoms **199**, 255 (2003).

[44] T. Yoshida, A. Ino, T. Mizokawa, A. Fujimori, Y. Taguchi, T. Katsufuji, and Y. Tokura, Photoemission spectral weight transfer and mass renormalization in the Fermi-liquid system La$_{1-x}$Sr$_x$TiO$_{3+y/2}$, EPL **59**, 258 (2002).

[45] M. Imada, A. Fujimori, and Y. Tokura, Metal-insulator transitions, Rev. Mod. Phys. **70**, 1039 (1998).

[46] A. Georges, G. Kotliar, W. Krauth, and M. J. Rozenberg, Dynamical mean-field theory of strongly correlated fermion systems and the limit of infinite dimensions, Rev. Mod. Phys. **68**, 13 (1996).

[47] G. Kotliar, S. Y. Savrasov, K. Haule, V. S. Oudovenko, O. Parcollet, and C. A. Marianetti, Electronic structure calculations with dynamical mean-field theory, Rev. Mod. Phys. **78**, 865 (2006).

[48] E. Pavarini, A. Yamasaki, J. Nuss, and O. K. Andersen, How chemistry controls electron localization in 3d$^1$ perovskites: a Wannier-function study, New J. Phys. **7**, 188 (2005).

[49] K. Yoshimatsu, K. Horiba, H. Kumigashira, T. Yoshida, A. Fujimori, and M. Oshima, Metallic Quantum Well States in Artificial Structures of Strongly Correlated Oxide, Science **333**, 319 (2011).

[50] A. Chikamatsu, H. Wadati, H. Kumigashira, M. Oshima, A. Fujimori, M. Lippmaa, K. Ono, M. Kawasaki, and H. Koinuma, Gradual disappearance of the Fermi surface near the metal-insulator transition in La$_{1-x}$Sr$_x$MnO$_3$ thin films, Phys. Rev. B **76**, 201103(R) (2007).

[51] T. Yoshida, X. J. Zhou, T. Sasagawa, W. L. Yang, P. V. Bogdanov, A. Lanzara, Z. Hussain, T. Mizokawa, A. Fujimori, H. Eisaki, Z. X. Shen, T. Kakeshita, and S. Uchida, Metallic Behavior of Lightly Doped La$_{2-x}$Sr$_x$CuO$_4$ with a Fermi Surface Forming an Arc, Phys. Rev. Lett. **91**, 027001 (2003).

[52] K. M. Shen, F. Ronning, D. H. Lu, W. S. Lee, N. J. C. Ingle, W. Meevasana, F. Baumberger, A. Damascelli, N. P. Armitage, L. L. Miller, Y. Kohsaka, M. Azuma, M. Takano, H. Takagi, and Z. X. Shen, Missing Quasiparticles and the Chemical Potential Puzzle in the Doping Evolution of the Cuprate Superconductors, Phys. Rev. Lett. **93**, 267002 (2004).

[53] K. M. Shen, F. Ronning, D. H. Lu, F. Baumberger, N. J. C. Ingle, W. S. Lee, W. Meevasana, Y. Kohsaka, M. Azuma, M. Takano, H. Takagi, and Z. X. Shen, Nodal Quasiparticles and Antinodal Charge Ordering in $Ca_{2-x}Na_xCuO_2Cl_2$, Science **307**, 901 (2005).

[54] W. F. Brinkman and T. M. Rice, Application of Gutzwiller's Variational Method to the Metal-Insulator Transition, Phys. Rev. B **2**, 4302 (1970).

[55] I. A. Nekrasov, K. Held, G. Keller, D. E. Kondakov, Th. Pruschke, M. Kollar, O. K. Andersen, V. I. Anisimov, and D. Vollhardt, Momentum-resolved spectral functions of $SrVO_3$ calculated by LDA+DMFT, Phys. Rev. B **73**, 155112 (2006).

[56] T. Valla, A. V. Fedorov, P. D. Johnson, B. O. Wells, S. L. Hulbert, Q. Li, G. D. Gu, and N. Koshizuka, Evidence for Quantum Critical Behavior in the Optimally Doped Cuprate $Bi_2Sr_2CaCu_2O_{8+\delta}$, Science **285**, 2110 (1999).

[57] A. F. Ioffe and A. R. Regel, Non-Crystalline, Amorphous and Liquid Electronic Semiconductors, Prog. Semicond. **4**, 237 (1960).

[58] K. Hong, S.-H. Kim, Y.-J. Heo, and Y.-U. Kwon, Metal–insulator transitions of $SrTi_{1-x}V_xO_3$ solid solution system, Solid State Commun. **123**, 305 (2002).

[59] H. Tsuiki, K. Kitazawa, and K. Fueki, The Donor Level of $V^{4+}$ and the Metal-Nonmetal Transition in $SrTi_{1-x}V_xO_3$, Jpn. J. Appl. Phys. **22**, 590 (1983).

[60] H. D. Zhou and J. B. Goodenough, X-ray diffraction, magnetic, and transport study of lattice instabilities and metal-insulator transition in $CaV_{1-x}Ti_xO_3$ ($0 \leqslant x \leqslant 0.4$), Phys. Rev. B **69**, 245118 (2004).

[61] M. Kobayashi, K. Tanaka, A. Fujimori, S. Ray, and D. D. Sarma, Critical Test for Altshuler-Aronov Theory: Evolution of the Density of States Singularity in Double Perovskite $Sr_2FeMoO_6$ with Controlled Disorder, Phys. Rev. Lett. **98**, 246401 (2007).

[62] K. W. Kim, J. S. Lee, T. W. Noh, S. R. Lee, and K. Char, Metal-insulator transition in a disordered and correlated $SrTi_{1-x}Ru_xO_3$ system: Changes in transport properties, optical spectra, and electronic structure, Phys. Rev. B **71**, 125104 (2005).

[63] J. Kim, J.-Y. Kim, B.-G. Park, and S.-J. Oh, Photoemission and x-ray absorption study of the electronic structure of $SrRu_{1-x}Ti_xO_3$, Phys. Rev. B **73**, 235109 (2006).

[64] D. D. Sarma, O. Rader, T. Kachel, A. Chainani, M. Mathew, K. Holldack, W. Gudat, and W. Eberhardt, Contrasting behavior of homovalent-substituted and hole-doped systems: O *K*-edge spectra from $LaNi_{1-x}M_xO_3$ (*M*=Mn, Fe, and Co) and $La_{1-x}Sr_xMnO_3$, Phys. Rev. B **49**, 14238 (1994).

# Supplemental Material

# Electronic structure of SrTi$_{1-x}$V$_x$O$_3$ films studied by *in situ* photoemission spectroscopy: Screening for a transparent electrode material

Tatsuhiko Kanda[1], Daisuke Shiga[1,2], Ryu Yukawa[2,†], Naoto Hasegawa[1], Duy Khanh Nguyen[1], Xianglin Cheng[1], Ryosuke Tokunaga[1], Miho Kitamura[2], Koji Horiba[2], Kohei Yoshimatsu[1], and Hiroshi Kumigashira[1,2,*]

[1] *Institute of Multidisciplinary Research for Advanced Materials (IMRAM), Tohoku University, Sendai, 980–8577, Japan*

[2] *Photon Factory, Institute of Materials Structure Science, High Energy Accelerator Research Organization (KEK), Tsukuba, 305–0801, Japan*

[†] Present Address: *Department of Applied Physics, Osaka University, Suita, Osaka 565-0871, Japan*

[*]Author to whom correspondence should be addressed: kumigashira@tohoku.ac.jp

## 1. Geometry in the ARPES measurements

Figure S1(a) shows a sketch of the experimental geometry for our *in situ* ARPES measurements. The incident synchrotron radiation beam and outgoing photoelectrons entering the analyzer slit define the emission plane, which is horizontal in this case. Light polarization (horizontal or vertical) refers to this emission plane. The reciprocal space along the X- and Y-axis direction is explored by varying the angles of $\theta$ and $\varphi$, respectively.

Figures S1(b) and S1(d) show the ARPES intensity maps along the Γ–X direction for a SrVO$_3$ film with 40 nm thickness with linear horizontal (LH) and linear vertical (LV) polarizations, respectively, compared with the corresponding results of the tight-binding calculation in the mass renormalization scheme for each 3$d$ $t_{2g}$ orbital (Fig. S1(c)). A significant polarization dependence can clearly be observed. Owing to the dipole selection rules for the present experimental setup, the ARPES intensities of the $d_{zx}$-derived bands are dominant in the LH mode (Fig. S1(b)), while those of the $d_{xy}$- and $d_{yz}$-derived bands are dominant in the LV mode (Fig. S1(d)). In other words, the polarization-dependent ARPES enabled the determination of the bands of the $d_{zx}$ and $d_{xy}/d_{yz}$ states separately.

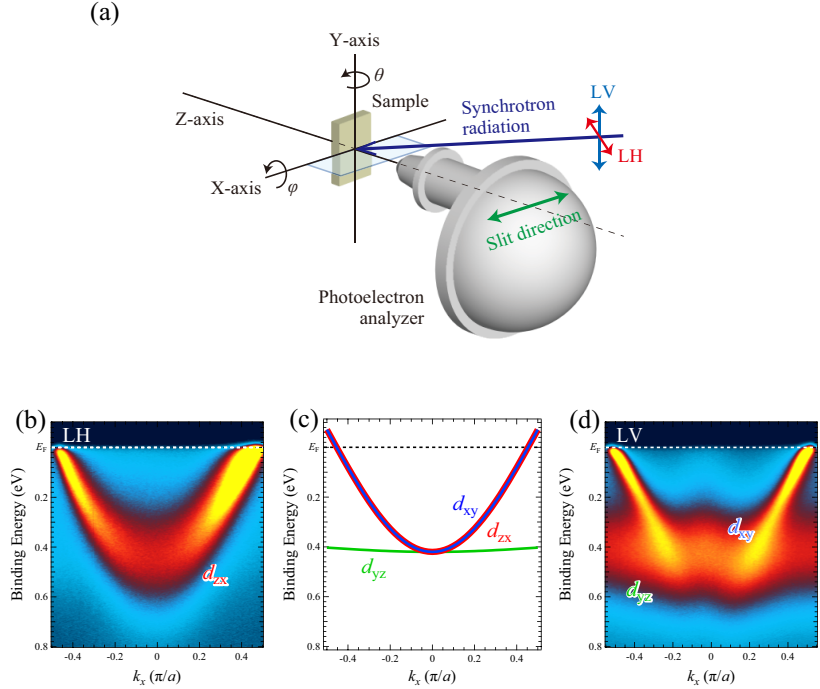

**FIG. S1:** (a) Sketch of our experimental geometry at the beamline BL-2A MUSASHI of Photon Factory, KEK. LH and LV are linear-horizontal and linear-vertical polarizations, respectively. In this geometry, the incoming photon momentum is coplanar to the entrance slit (photoelectron

S2

detection) direction of the photoelectron analyzer.   (b,d) ARPES intensity maps at a photon energy of 88 eV along the Γ–X direction for an SrVO$_3$ film with LH (b) and LV (d) lights.   (c) Tight-binding calculation result for the SrVO$_3$ films.   The blue, green, and red curves represent the $d_{xy}$, $d_{yz}$, and $d_{zx}$ bands, respectively.   It is evident that the ARPES intensity of the $d_{zx}$- ($d_{yz}$/$d_{xy}$-) derived bands becomes dominant in the LH (LV) mode.

## 2. Characterization of STVO films

### 2.1 Surface quality of measured STVO films

The surface morphologies of the prepared SrTi$_{1-x}$V$_x$O$_3$ (STVO) films were also analyzed using atomic force microscopy (AFM), as shown in Fig. S2, together with the RHEED images. Atomically flat surfaces with step-and-terrace structures were clearly observed for all samples, indicating that the measured surfaces were atomically flat.

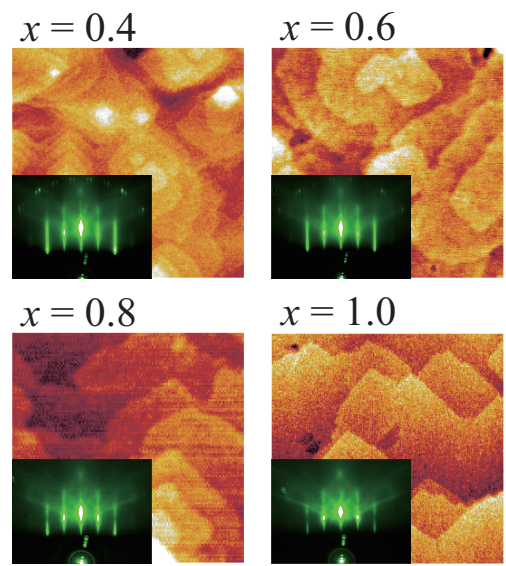

FIG S2: AFM images of STVO films (scan area: 0.5 x 0.5 μm$^2$).   RHEED patterns taken immediately after the growth of each film are also shown in the insets.

### 2.2 Crystallinity of STVO films

Figure S3(a) shows the XRD pattern around the (002) diffraction peaks of the STVO film. As shown in Fig. S3(a), the out-of-plane XRD scans confirm a single-phase STVO solid solution over the entire composition range of $0.4 \leq x \leq 1.0$. Furthermore, a sharp diffraction pattern with well-defined Laue fringes can be clearly observed, indicating the high quality of the films, i.e., homogeneously coherent films with atomically flat surfaces and chemically abrupt interfaces. Because SrVO$_3$ (SVO) has a smaller lattice parameter ($a_{SVO}$ = 3.843 Å [34]) than SrTiO$_3$ (STO) ($a_{STO}$ = 3.905 Å [35]), the (002) peak of STVO systematically approaches that of the STO substrate with decreasing $x$, reflecting a decrease in the mismatch between the films and substrates [36]. This behavior is further confirmed in the plot of $c$-axis length vs. $x$ shown in Fig. 3(b). The almost linearly proportional relationship indicates that the lattice of STVO obeys Vegard's law.

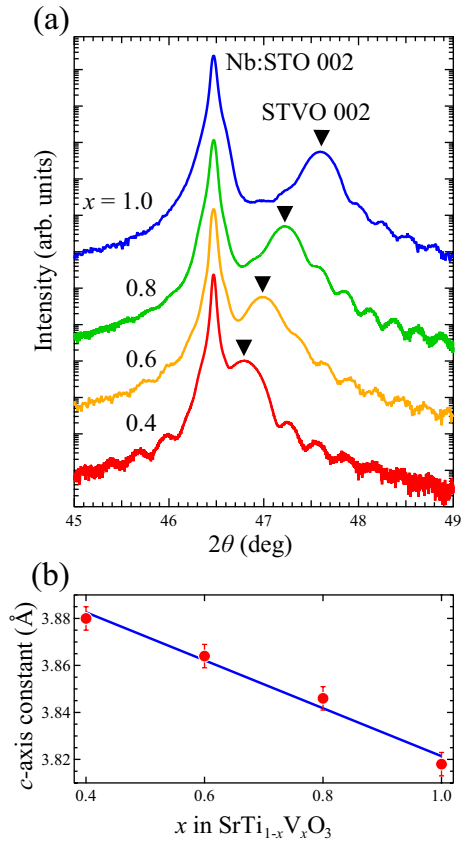

FIG. S3: (a) XRD pattern along the (00$L$) scan direction around the (002) reflections of STVO films and Nb:SrTiO$_3$ substrates. (b) Plot of $c$-axis parameter as a function of $x$.

## 2.3 Resistivity of STVO films

Figure S4 shows the electrical resistivity vs. temperature ($\rho$–$T$) curves for STVO films grown on (LaAlO$_3$)$_{0.3}$−(SrAl$_{0.5}$Ta$_{0.5}$O$_3$)$_{0.7}$ (LSAT) substrates under almost the same growth conditions as the SrTiO$_3$ substrates. It should be noted that the transport properties are in good agreement with those in a previous report on STVO films grown on LSAT [36]. With decreasing $x$, the resistivity over the entire temperature range increases. For $x \geq 0.6$, the resistivity of the films shows a clear metallic behavior, i.e., a monotonic increase with increasing temperature, although the $\rho$–$T$ for $x = 0.6$ shows an upturn at low temperatures owing to weak localization. In contrast, for $x = 0.4$, the resistivity monotonically decreases with increasing temperature, indicating the insulating nature of STVO at $x = 0.4$ due to strong localization. Thus, these results suggest the occurrence of composition-driven MIT at $x = 0.4$–0.6, which is consistent with the photoemission results.

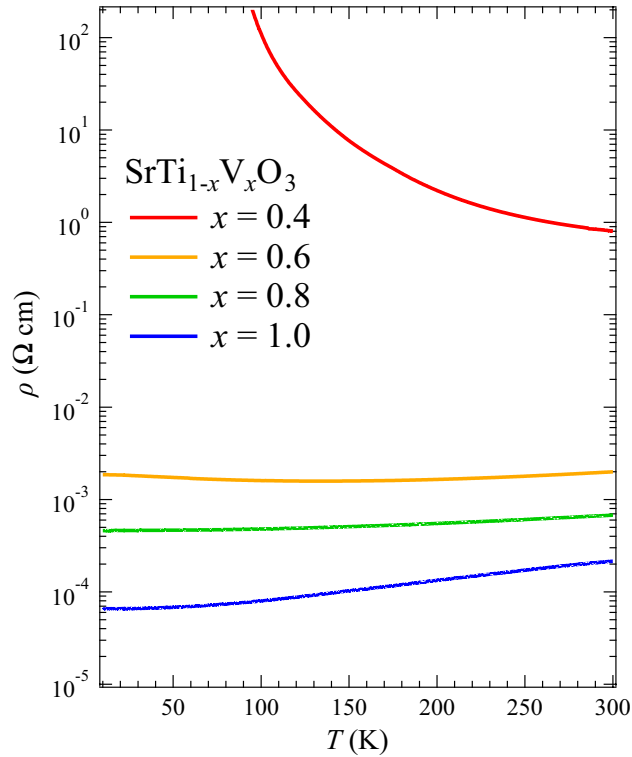

**FIG S4:** Temperature-dependent resistivity of STVO films with variable $x$. The samples for the resistivity measurements were grown on LSAT substrates to prevent electric current from flowing through the conducting substrate. Note that the composition dependence of the $\rho$–$T$ curves is in good agreement with a previous report [36].

### 3. Fermi surface of SrVO$_3$

The Fermi surface (FS) topology of bulk SrVO$_3$ (SVO) is illustrated in Fig. S5(a). The FS of SrVO$_3$ consists of three cylindrical FS sheets originating from three V 3$d$ $t_{2g}$ states: $d_{xy}$ (blue), $d_{yz}$ (green), and $d_{zx}$ (red) [48].   In the ARPES measurements obtained at a photon energy of 88 eV, the ARPES spectra almost trace the band dispersion in the ΓXM emission plane ($k_x$–$k_y$ plane with $k_z$ = 0, as indicated by the yellow plate) of the Brillouin zone, where $k_x$ and $k_y$ are the momenta parallel to the film surface and $k_z$ is the momentum perpendicular to the surface [23,49]. Consequently, in the measurement plane (on the yellow plate), a circular FS centered at the Γ point was derived from the $d_{xy}$ band, while the two nearly parallel line FSs are from the $d_{yz}$ and $d_{zx}$ bands [Fig. S5(b)].   The characteristic FS of SrVO$_3$ can be clearly in the FS mapping results shown in Fig. S5(c).

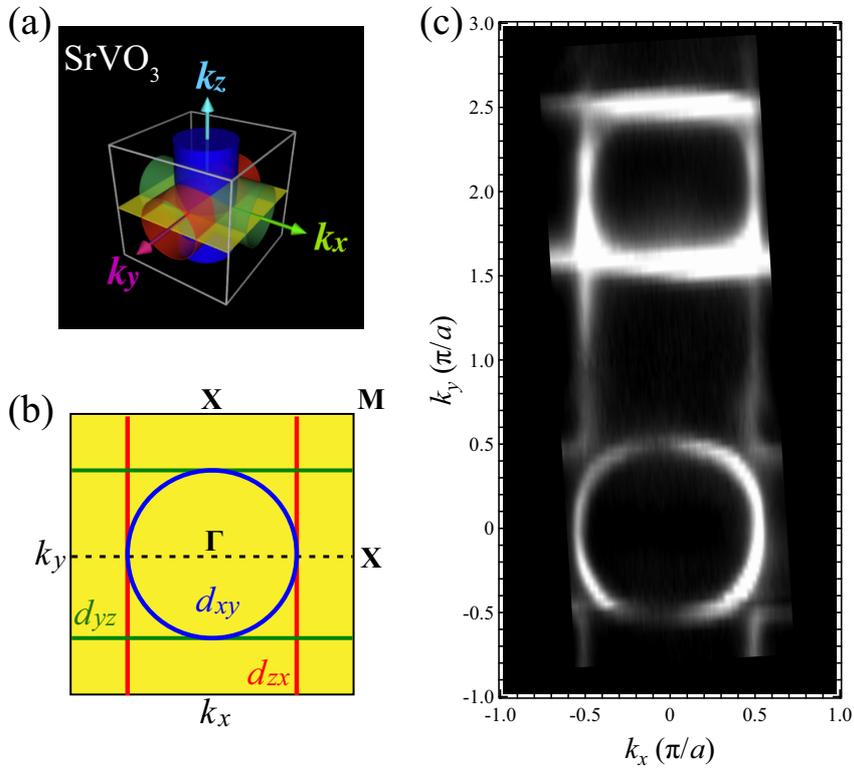

**FIG. S5:** (a) Schematic illustration of the FS topology of SrVO$_3$ and (b) FS at the ΓXM emission plane.   The ARPES images shown in the main text were measured along the Γ–X high-symmetry line ($k_x$ slice at $k_y$ = 0), as indicated by the dashed line.   (c) Results of FS mapping for SrVO$_3$ films.   Note that the FS mapping was carried out using circularly polarized light to average the polarization effects.

## 4. Analysis of ARPES images

Figure S6 shows the selected ARPES spectra energy distribution curves (EDCs) from the ARPES images of the STVO films. Parabolic dispersive features corresponding to the $d_{zx}$ bands can be clearly observed in the ARPES spectra, especially for the SVO film. The intensity modulation of the ARPES peaks is due to pronounced matrix element effects with respect to the Γ point, which can be clearly seen in the ARPES spectral weight plots in the energy–momentum ($E$–$k$) space shown in Fig. S7. The quasiparticle spectral weight near the Fermi level ($E_F$) significantly weakens with decreasing $x$ and eventually almost completely smears out at $x = 0.4$.

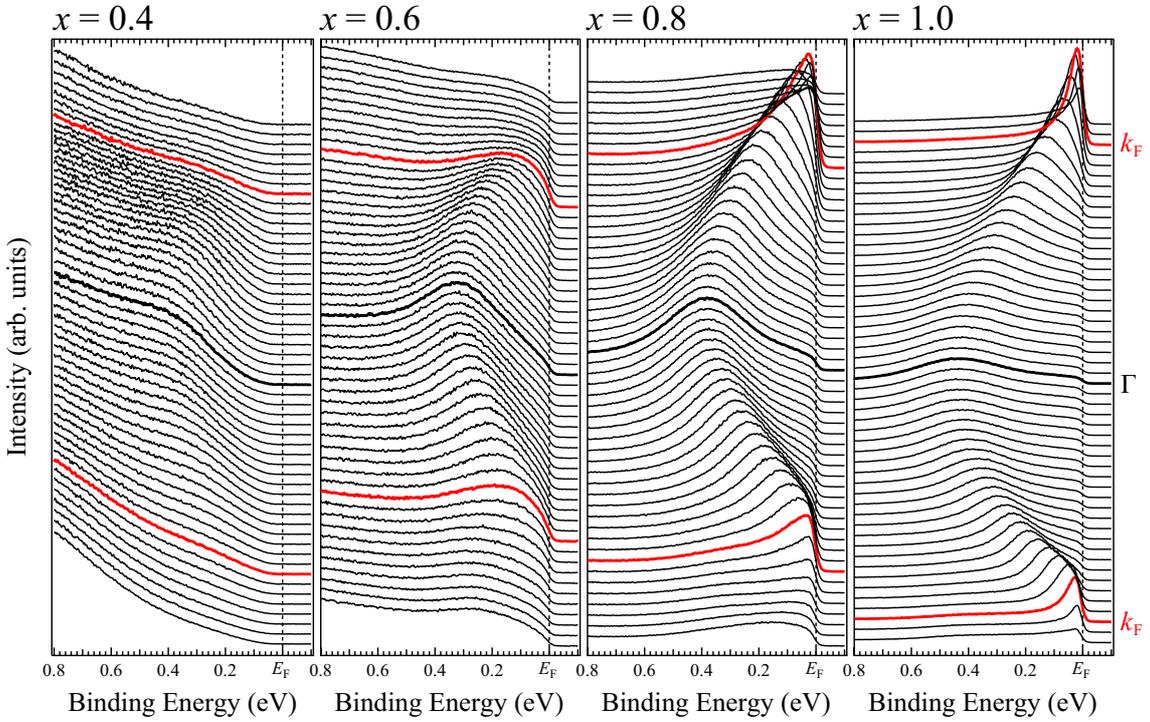

**FIG. S6:** ARPES spectra of STVO films.

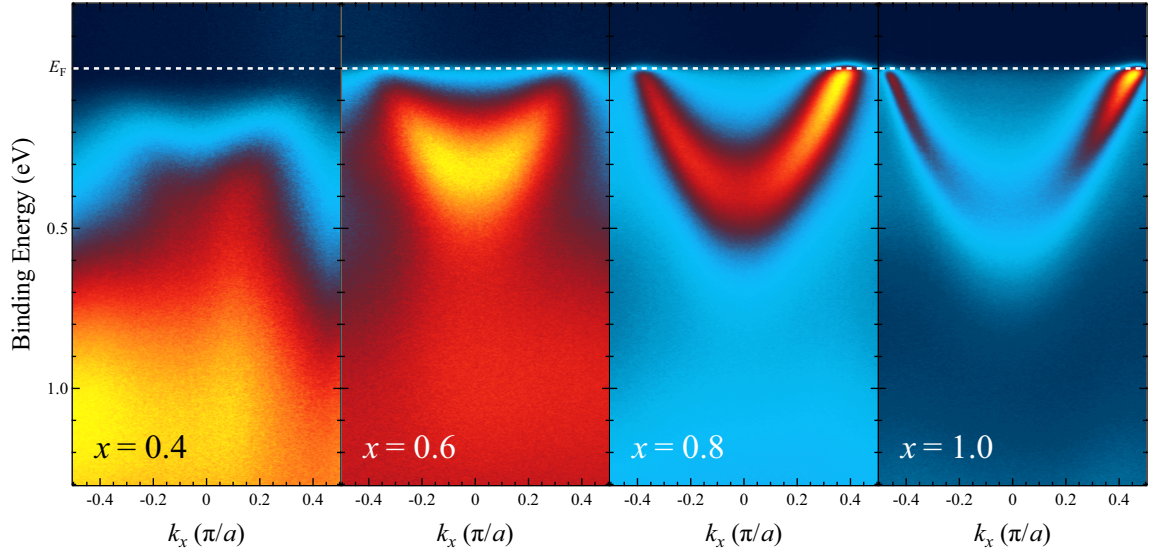

**FIG. S7**: ARPES images for STVO films in $E$–$k_x$ space. The intensity modulation with respect to the Γ point is due to pronounced matrix element effects in the ARPES measurements. Note that the ARPES images in Fig. 3 in the main text are symmetrized with respect to the center line ($k_x = 0$) and averaged.